\documentclass[twocolumn,showpacs,%
aps,superscriptaddress,
prd,notitlepage,showkeys,
nofootinbib]{revtex4-1}

\usepackage{ulem}
\usepackage{amssymb}
\usepackage{amsmath}
\usepackage{graphicx}
\usepackage{dcolumn}
\usepackage[colorlinks,urlcolor=blue,citecolor=blue,linkcolor=blue]{hyperref}
\usepackage{color,units}
\usepackage[dvipsnames]{xcolor} 
\usepackage{lineno}
\usepackage{xspace}
\usepackage{longtable} 
\usepackage{float} 
\usepackage{multirow}
\usepackage{amsfonts,wasysym,epsfig,verbatim,subfigure,bm,mathrsfs,lipsum}
\begin{document}
\newcommand{\IUCAA}{Inter-University Centre for Astronomy and
Astrophysics, Post Bag 4, Ganeshkhind, Pune 411 007, India}
\newcommand{\OKC}{The Oskar Klein Centre, Department of Astronomy, Stockholm University, AlbaNova, SE-10691 Stockholm,
Sweden}

\newcommand{\MPI}{Max-Planck-Institut f{\"u}r Gravitationsphysik (Albert-Einstein-Institut), D-30167 Hannover, Germany}

\newcommand{\LBNZ}{Leibniz Universit{\"a}t Hannover, D-30167 Hannover, Germany}

\newcommand{\sayak}[1]{{\color{red}\bf  SD:} {\color{red} #1}}

\newcommand{\bhaskar}[1]{{\color{blue}\bf  BB:} {\color{blue} #1}}
\title{Constraining neutron star properties with a new equation of state insensitive approach}
\author{Bhaskar Biswas}\email{phybhaskar95@gmail.com}
\affiliation{\IUCAA}\affiliation{\OKC}

\author{Sayak Datta}\email{sayak.datta@aei.mpg.de}
\affiliation{\IUCAA}\affiliation{\MPI}\affiliation{\LBNZ}
\date{\today}

\begin{abstract}
Instead of parameterizing the pressure-density relation of a neutron star (NS), one can parameterize its macroscopic properties such as mass ($M$), radius ($R$), and dimensionless tidal deformability ($\Lambda$) to infer the equation of state (EoS) combining electromagnetic and gravitational wave (GW) observations. We present a new method to parameterize $R(M)$ and $\Lambda(M)$ relations, which approximate the candidate EoSs with  accuracy better than 5\% for all masses and span a broad region of $M-R-\Lambda$ plane. Using this method we combine the $M-\Lambda$ measurement from GW170817 and GW190425, and simultaneous $M-R$ measurement of PSR J0030+0451 and PSR J0740+6620 to place joint constraints on NS properties. 
At 90 \% confidence, we infer $R_{1.4}=12.05_{-0.87}^{+0.98}$ km and $\Lambda_{1.4}=372_{-150}^{+220}$ for a $1.4 M_{\odot}$ NS, and $R_{2.08}=12.65_{-1.46}^{+1.36}$ km for a $2.08 M_{\odot}$  NS. Furthermore, we use the inferred values of the maximum mass of a nonrotating NS $M_{\rm max}=2.52_{-0.29}^{+0.33} M_{\odot}$ to investigate the nature of the secondary objects in three potential neutron star-black hole merger (NSBH) system.
\end{abstract}

\maketitle

\section{Introduction}
The lack of theoretical understanding of the matter properties of neutron star (NS) at the supra-nuclear densities has prevented us to give a unique description of its EoS. The recent detection of gravitational waves (GWs) by LIGO~\cite{advanced-ligo}/Virgo~\cite{advanced-virgo} (LVC) collaboration form two likely binary neutron star (BNS) merger event named GW170817~\cite{TheLIGOScientific:2017qsa,Abbott:2018exr} and GW190425~\cite{Abbott:2020uma} has ushered in a new light to solve this problem. The tidally deformed components of a BNS merger in their late inspiral phase leave a detectable imprint on the emitted GW signals. The measurement of the tidal deformabilities associated with these observations provide direct information about the equation of state (EoS) of the NS~\cite{Flanagan:2007ix,Hinderer:2007mb,Damour:2009vw,Binnington:2009bb}. It also depends on the superfluid nature of the matter, and symmetry of space time~\cite{Char:2018grw, Datta:2019ueq, Biswas:2019gkw,Biswas:2019ifs}. Apart from GWs, other type of astrophysical observations also have supplied complementary constraints on the properties of NS. Space-based mission led by Neutron Star Interior Composition ExploreR (NICER) collaboration have already presented quite precise measurement of mass and radius of PSR J0030+0451~\cite{Miller:2019cac,Riley:2019yda} and PSR J0740+6620~\cite{Miller:2021qha,Riley:2021pdl} using pulse-profile modeling. Additionally, observations of massive pulsars~\cite{Antoniadis:2013pzd,Cromartie:2019kug,2021arXiv210400880F} by radio telescopes provide important insights on the high density part of the NS EoS. All of these measurements of NS macroscopic properties, such as mass ($M$), radius ($R$), and tidal deformablities ($\lambda$), lead us to the EoS which is the same for all the NSs present in the universe.

To maximize the amount of information, we need to combine all the present and future observations from different astrophysical messengers. Several parametric~\cite{Raaijmakers:2019dks,Jiang:2019rcw,Most:2018hfd,Traversi:2020aaa,Xie:2019sqb,Biswas_arXiv_2008.01582B,Al-Mamun:2020vzu,Dietrich:2020efo,Biswas:2021yge,Miller:2021qha} or nonparametric~\cite{Landry_2020PhRvD.101l3007L,han2021bayesian} approaches already exist in the literature which successfully combine information of NS properties from multiple cosmic messengers and place joint constraints on its EoS. All the parametric approaches construct a particular functional form of NS EoS which span a wide range in the pressure-density ($P-\rho$) plane maintaining thermodynamic stability and  causality. Since the EoS has one-to-one correspondence with $R (M)$ or $\Lambda (M)$ (where $\Lambda = \lambda/M^5$), by choosing $P(\rho)$ parameterization we can reconstruct the posterior of NS EoS adopting a hierarchical Bayesian formalism. Though parameterizing $P(\rho)$ relation gives us several advantages as it tells us about the microphysics, however computationally it is expensive. One needs to solve both Tolman–Oppenheimer–Volkoff (TOV) equation~\cite{TOV} and perturbed Einstein equations to calculate $R (M)$ and $\Lambda (M)$ for a given EoS and this process needs to be repeated over a million times to ensure well converged posterior distribution of EoS parameters. In the nonparametric method, a large number of EoS functionals are generated whose ranges in the $P-\rho$
plane are loosely guided by a certain number of widely
used candidate nuclear EoSs from the literature, without
the explicit need for any type of parameterization. Using such method $P-\rho$ plane can be constructed which can be used to compute the macroscopic properties of NSs by solving TOV equation in the similar manner as the parametric methods.

However, instead of parameterizing $P(\rho)$ relation, one can choose to parameterize $R (M)$ and $\Lambda (M)$ relation. In past, Refs.~\cite{DelPozzo:2013ala,Agathos_2015} have considered a simple expansion of $\lambda (M)$ about a reference mass of $1.4 M_{\odot}$: $\lambda (M) = \sum_{j}\frac{1}{j!} c_j (\frac{M-1.4M_{\odot}}{M_{\odot}})^j$. The set of expansion coefficients $\{c_j\}$ is EoS-sensitive and can be used to approximate different EoSs. However, because of the simplicity of this model this approximation breaks down for most of the EoSs above $1.8 M_{\odot}$~\cite{Chatziioannou:2020pqz}. Also for a set of $\{c_j\}$ it is not possible to calculate the maximum mass ($M_{\rm max}$) of a NS in terms of $\{c_j\}$. Therefore the constraints coming from the massive pulsar observations cannot be added under this framework using Bayesian statistics since the probability is associated with $M_{\rm max}$. Nonetheless, a better parameterization of $R (M)$ and $\Lambda (M)$ relation overcoming the above-mentioned difficulties would be very useful to speed up the parameter estimation combining multiple observations. In the parametric or the nonparametric approaches, the dominant cost in parameter estimation comes from evaluating $R(M),\Lambda(M)$ for each EoS associated with the corresponding observed masses. By directly parameterizing $R (M)$ and $\Lambda (M)$ relation, we can avoid this step and reduce a huge computational cost.

In this paper, we introduce a new way to parameterize $R (M)$ and $\Lambda (M)$ relations which approximate the candidate EoSs quite accurately for all masses. We then apply this parameterization to combine $M-\Lambda$ measurements from GW170817 and GW190425 observations, $M-R$ measurements of PSR J0030+0451 and PSR J0740+6620, and massive pulsar mass measurements to place joint constraints on the properties of NS. Finally based on the inferred $M_{\rm max}$, we investigate the nature of the secondary objects in three potential neutron star-black hole (NSBH) systems announced by LVC in their third observing run.

\section{Usage of $M-\Lambda$ curve}
We use the following functional form to parameterize the $M-\Lambda$ curve,
\begin{equation}
    \frac{M}{M_{\odot}}(\Lambda)=\sum_{i=0}^{3}b_i(\ln{\Lambda})^i \,.
\end{equation}

For a given set of $\{b_i\}$, the maximum mass ($M_{\rm max}$) is defined as, 
\begin{equation}
     \left.\frac{dM}{d \Lambda}\right\vert_{M = M_{\rm max}} = 0\,.
     \label{eq:mmax}
\end{equation}

Therefore, $M_{\rm max} = \sum_{i=0}^{3}b_i(\ln{\Lambda_{\rm max}})^i$, where $\ln{\Lambda_{\rm max}} = \mathrm{min} (-b_2\pm \sqrt{b_2^2-3b_1b_3})/3b_3$. We impose following conditions on this parameterization: (i) $b_2^2-3b_1b_3 > 0$, otherwise $M_{\rm max}$ is imaginary. (ii) $\Lambda (M = 1 M_{\odot}) < 5000$, which takes into account of the range covered by the candidate EoSs.

For this work, we consider 27 EoS models which are computed from different nuclear-physics approximations covering a wide-range in mass-radius diagram (See Fig. 1 of~\cite{Biswas:2021pvm}).  These EoSs are taken from publicly available~{\tt LalSuite}~\citep{lalsuite} package. They are consisted of plain $npe\mu$ nuclear matter which include--- (i) Variational-method EoSs (AP3-4) and APR~\citep{APR4_EFP}, APR4\_EFP~\citep{Endrizzi_2016,APR4_EFP}, WFF1-2~\citep{wff1}), (ii) potential based EoS SLY~\citep{Douchin:2001sv}, (iii) nonrelativistic Skyrme interactions based EoS (SLY2 and SLY9~\citep{SLy2_1,SLy2_2}, SLY230A~\citep{SLy230A_3}, RS~\citep{RS_3}, BSK20 and BSK21~\citep{Goriely_2010,Pearson_2011}, SK255 and SK272~\citep{SLy2_1,SLy2_2,SK255_3}, SKI2-6~\citep{SLy2_1,SLy2_2,REINHARD1995467}, SKMP~\citep{SLy2_1,SLy2_2,SKMP_3}), (iv) relativistic Brueckner-Hartree-Fock EOSs (MPA1~\citep{MPA1}, ENG~\citep{Engvik_1994}), (v) relativistic mean field theory EoSs (MS1, MS1B, MS1\_PP, MS1B\_PP where MS1\_PP, MS1B\_PP~\citep{MS_PP} are the analytic piecewise polytrope fits of original MS1 and MS1B EoS, respectively). It should be noted that for the choice of EoS catalog we closely follow Ref.~\cite{Biswas:2021pvm,LIGOScientific:2019eut} but excluding EoSs with phase transition. These EoSs are compatible with the mass measurement ($M =2.08 \pm 0.07 M_{\odot}$ at $1 \sigma$ confidence interval) of the observed heaviest pulsar~\citep{Cromartie:2019kug,2021arXiv210400880F}. In Table 1 of Ref.~\cite{Biswas:2021pvm}, the radius and tidal deformability of $1.4 M_{\odot}$ NS are shown for each EoS and the corresponding ranges are (10.42, 15.07) km and (153,1622) respectively. Readers are referred to the respective references listed in Table~\ref{tab:best-fit} and as well as Ref.~\citep{LIGOScientific:2019eut} for the details on these EoSs. 

\begin{table*}[ht!]

\begin{tabular}{|p{3cm}|p{1.5cm}|p{1.5cm}|p{1.5cm}|p{1.5cm}|p{1.5cm}|p{1.5cm}|p{1.5cm}|p{1.5cm}|}
\toprule
EoS name &    $b_0$ &    $b_1$ &     $b_2$ &    $b_3$ & maximum \% error in $M (\Lambda)$ fit & true $M_{\rm max} [M_{\odot}]$ & fitted $M_{\rm max} [M_{\odot}]$ & \% error in $M_{\rm max}$ \\
\hline
      AP3~\cite{APR4_EFP} &  2.433770 &  0.022338 &  -0.054348 &  0.003612 &                                2.317 &                           2.390 &                              2.436 &                1.891 \\
      AP4~\cite{APR4_EFP} &  2.242567 &  0.031303 &  -0.052220 &  0.003488 &                                1.716 &                           2.213 &                              2.247 &                1.538 \\
 APR4\_EPP~\cite{Endrizzi_2016,APR4_EFP} &  2.218288 &  0.031008 &  -0.051365 &  0.003410 &                                1.689 &                           2.159 &                              2.223 &                2.874 \\
    BSK20~\cite{Goriely_2010,Pearson_2011} &  2.168145 &  0.075493 &  -0.056412 &  0.003518 &                                1.565 &                           2.167 &                              2.195 &                1.246 \\
    BSK21~\cite{Goriely_2010,Pearson_2011} &  2.251723 &  0.120568 &  -0.065303 &  0.003888 &                                1.791 &                           2.277 &                              2.311 &                1.469 \\
      ENG~\cite{Engvik_1994} &  2.238523 &  0.074432 &  -0.055334 &  0.003304 &                                0.359 &                           2.253 &                              2.265 &                0.520 \\
     MPA1~\cite{MPA1} &  2.530978 &  0.033075 &  -0.058213 &  0.003810 &                                2.774 &                           2.469 &                              2.536 &                2.647 \\
     MS1B~\cite{MS_PP} &  2.918873 &  0.035826 &  -0.060997 &  0.003727 &                                4.531 &                           2.797 &                              2.924 &                4.347 \\
  MS1B\_PP~\cite{MS_PP} &  2.819530 &  0.058321 &  -0.063682 &  0.003841 &                                2.339 &                           2.747 &                              2.833 &                3.062 \\
   MS1\_PP~\cite{MS_PP} &  2.838732 &  0.044147 &  -0.059261 &  0.003532 &                                2.570 &                           2.753 &                              2.847 &                3.313 \\
      MS1~\cite{MS_PP} &  2.939461 &  0.020344 &  -0.056119 &  0.003388 &                                4.878 &                           2.799 &                              2.941 &                4.838 \\
      SLY~\cite{Douchin:2001sv} &  2.025575 &  0.106152 &  -0.056934 &  0.003397 &                                1.258 &                           2.054 &                              2.078 &                1.187 \\
     WFF1~\cite{wff1} &  2.175612 &  0.007872 &  -0.050936 &  0.003711 &                                1.915 &                           2.137 &                              2.176 &                1.779 \\
     WFF2~\cite{wff1} &  2.237742 &  0.009126 &  -0.047259 &  0.003182 &                                1.558 &                           2.201 &                              2.238 &                1.653 \\
      APR~\cite{APR4_EFP} &  2.217008 &  0.033981 &  -0.052326 &  0.003495 &                                1.665 &                           2.190 &                              2.223 &                1.473 \\
       RS~\cite{SLy2_1,SLy2_2,RS_3} &  2.057707 &  0.135273 &  -0.057613 &  0.003175 &                                1.421 &                           2.117 &                              2.143 &                1.238 \\
    SK255~\cite{SLy2_1,SLy2_2,SK255_3} &  2.101598 &  0.123621 &  -0.057149 &  0.003211 &                                1.593 &                           2.144 &                              2.173 &                1.335 \\
    SK272~\cite{SLy2_1,SLy2_2,SK255_3} &  2.213105 &  0.107295 &  -0.057295 &  0.003281 &                                1.736 &                           2.232 &                              2.266 &                1.537 \\
     SKI2~\cite{SLy2_1,SLy2_2,REINHARD1995467} &  2.092382 &  0.144265 &  -0.058055 &  0.003108 &                                1.457 &                           2.163 &                              2.189 &                1.195 \\
     SKI3~\cite{SLy2_1,SLy2_2,REINHARD1995467} &  2.186424 &  0.136229 &  -0.060239 &  0.003311 &                                1.609 &                           2.240 &                              2.269 &                1.285 \\
     SKI4~\cite{SLy2_1,SLy2_2,REINHARD1995467} &  2.133640 &  0.122718 &  -0.061493 &  0.003598 &                                1.615 &                           2.170 &                              2.199 &                1.340 \\
     SKI5~\cite{SLy2_1,SLy2_2,REINHARD1995467} &  2.187534 &  0.130339 &  -0.055545 &  0.002922 &                                1.600 &                           2.240 &                              2.270 &                1.293 \\
     SKI6~\cite{SLy2_1,SLy2_2,REINHARD1995467} &  2.158231 &  0.120201 &  -0.061455 &  0.003598 &                                1.665 &                           2.190 &                              2.221 &                1.393 \\
     SKMP~\cite{SLy2_1,SLy2_2,SKMP_3} &  2.060798 &  0.125223 &  -0.058028 &  0.003301 &                                1.402 &                           2.107 &                              2.133 &                1.220 \\
     SLY2~\cite{SLy2_1,SLy2_2} &  2.022697 &  0.109539 &  -0.057869 &  0.003462 &                                1.298 &                           2.054 &                              2.078 &                1.163 \\
  SLY230A~\cite{SLy2_1,SLy2_2,SLy230A_3} &  2.074374 &  0.111313 &  -0.060514 &  0.003666 &                                1.553 &                           2.099 &                              2.129 &                1.395 \\
     SLY9~\cite{SLy2_1,SLy2_2} &  2.126594 &  0.116213 &  -0.060068 &  0.003528 &                                1.566 &                           2.156 &                              2.186 &                1.390 \\
 \hline
\end{tabular}

\caption{Parameters that provide the best fit to the candidate EoSs and corresponding maximum \% error are given. As well as true $M_{\rm max}$, fitted $M_{\rm max}$ obtained using Eq~\ref{eq:mmax}, and corresponding \% error in $M_{\rm max}$ for each candidate EoSs are quoted.}

\label{tab:best-fit}
\end{table*}


\begin{table}[ht]
\label{prior_tab}
    \centering
    \begin{tabular}{|c|c|}
         \hline
         Parameter& Prior  \\
         \hline
         $ b_0$ & uniform(2.007073 , 2.939462) \\
         \hline
         $ b_1$  & uniform(0.007872 , 0.165409)\\
         \hline
         $ b_2$  & uniform(-0.047259 , -0.077523)\\
         \hline
         $ b_3$  & uniform(0.002922 , 0.005051)\\
         \hline
    \end{tabular}
    \caption{Prior ranges of various EoS-sensitive parameters. }
\end{table}

\begin{figure}[ht]
    \centering
    \includegraphics[width=0.45\textwidth]{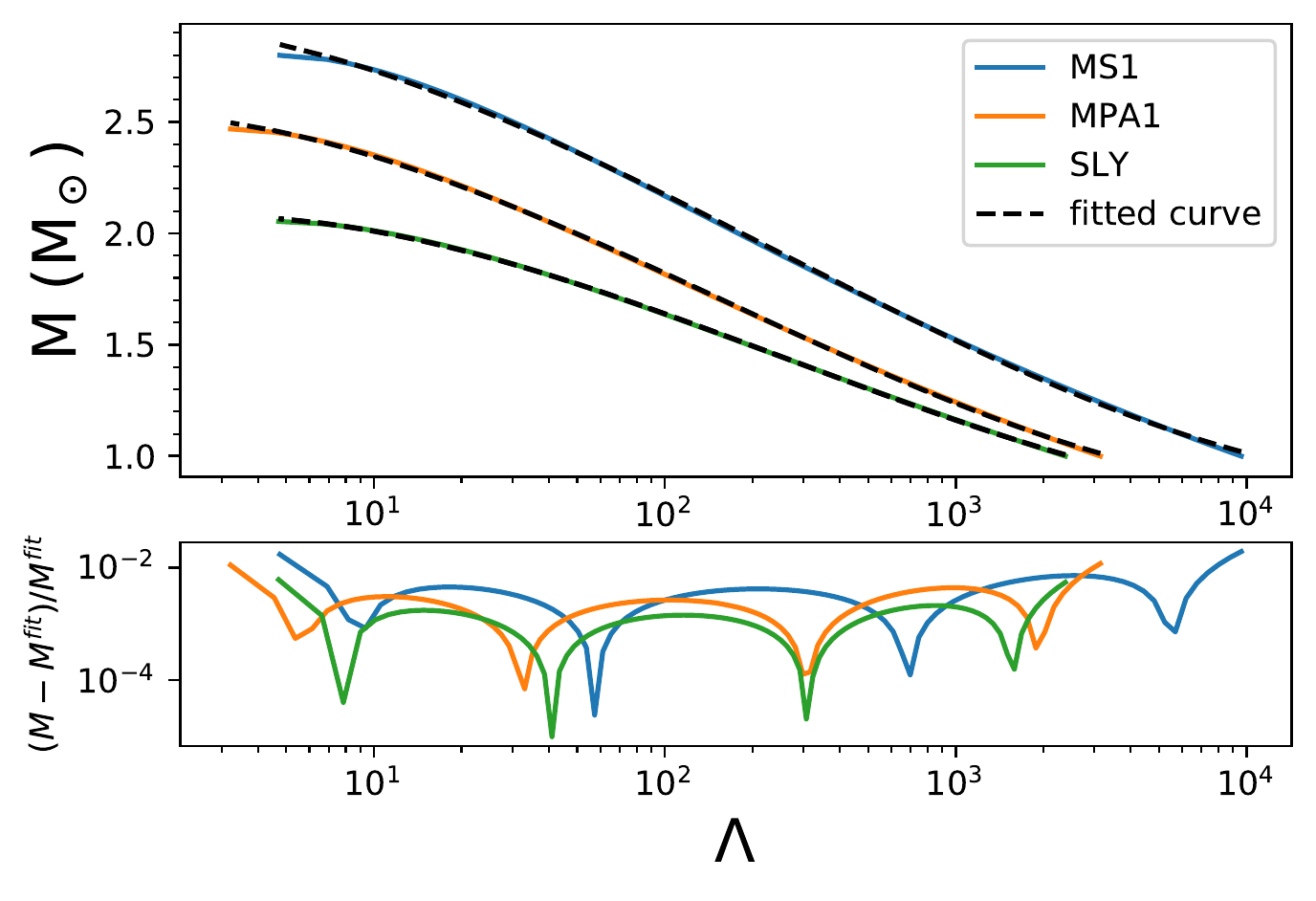}
    \caption{In upper panel, mass ($M$) and its fitted curve is plotted for three candidate EoSs SLy (soft)~\cite{Douchin:2001sv}, MPA1 (intermediate)~\cite{MPA1}, and MS1 (stiff)~\cite{MS_PP}  as a function of dimensionless tidal deformability ($\Lambda$). In lower panel, the percentage of error of the fit is also shown.}
    \label{fig:M-Lambda-fit}
\end{figure}

\begin{figure*}[ht!]
    \centering
    \includegraphics[width=\textwidth]{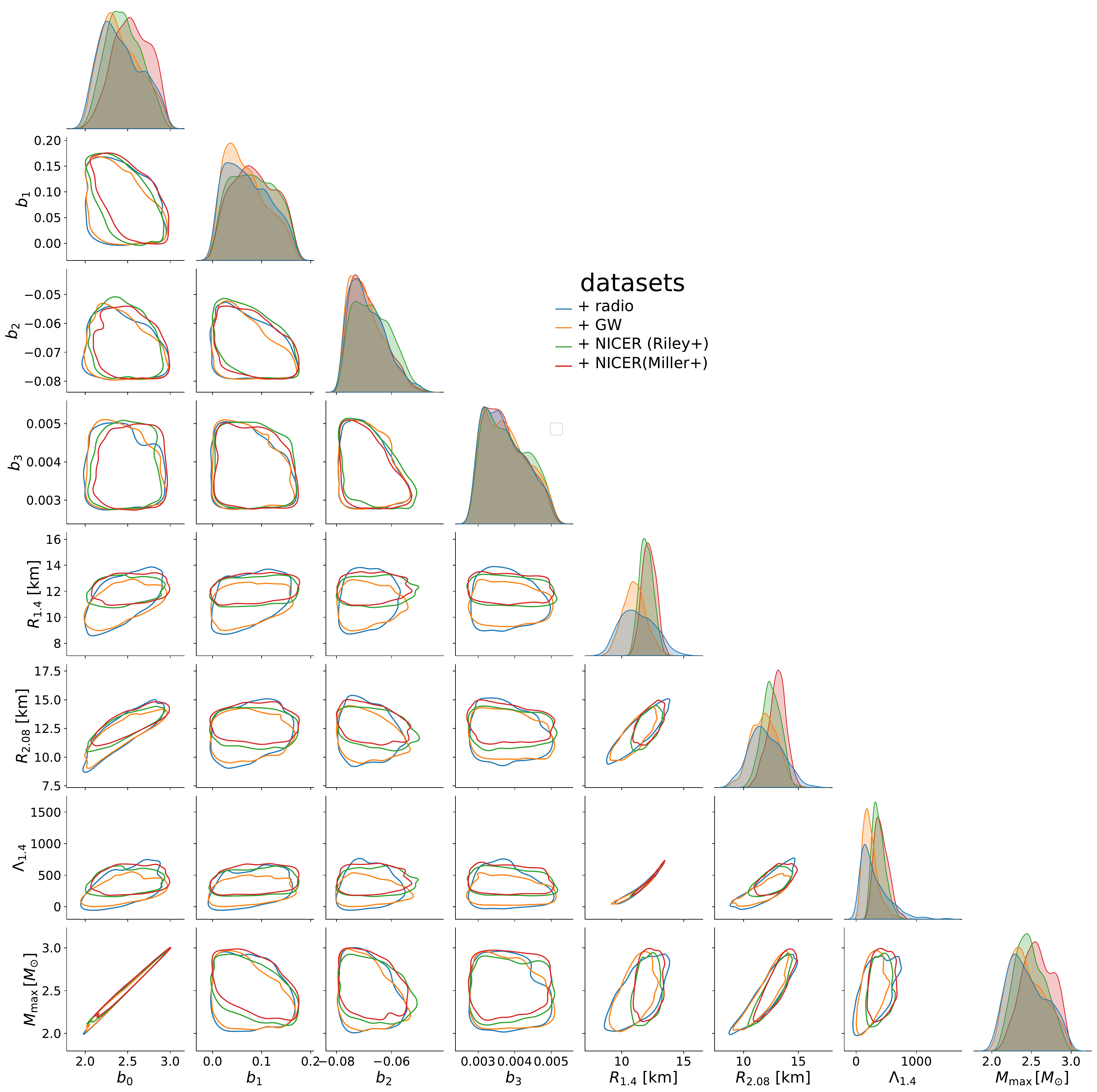}
    \caption{Posterior distribution of EoS-sensitive parameters and their correlations with $R_{1.4}$, $\Lambda_{1.4}$, $R_{2.08}$, and $M_{\rm max}$ are shown here after adding each type of observational data successively.}
    \label{fig:post-params}
\end{figure*}

\begin{figure*}[ht!]
    \centering
    \includegraphics[width=\textwidth]{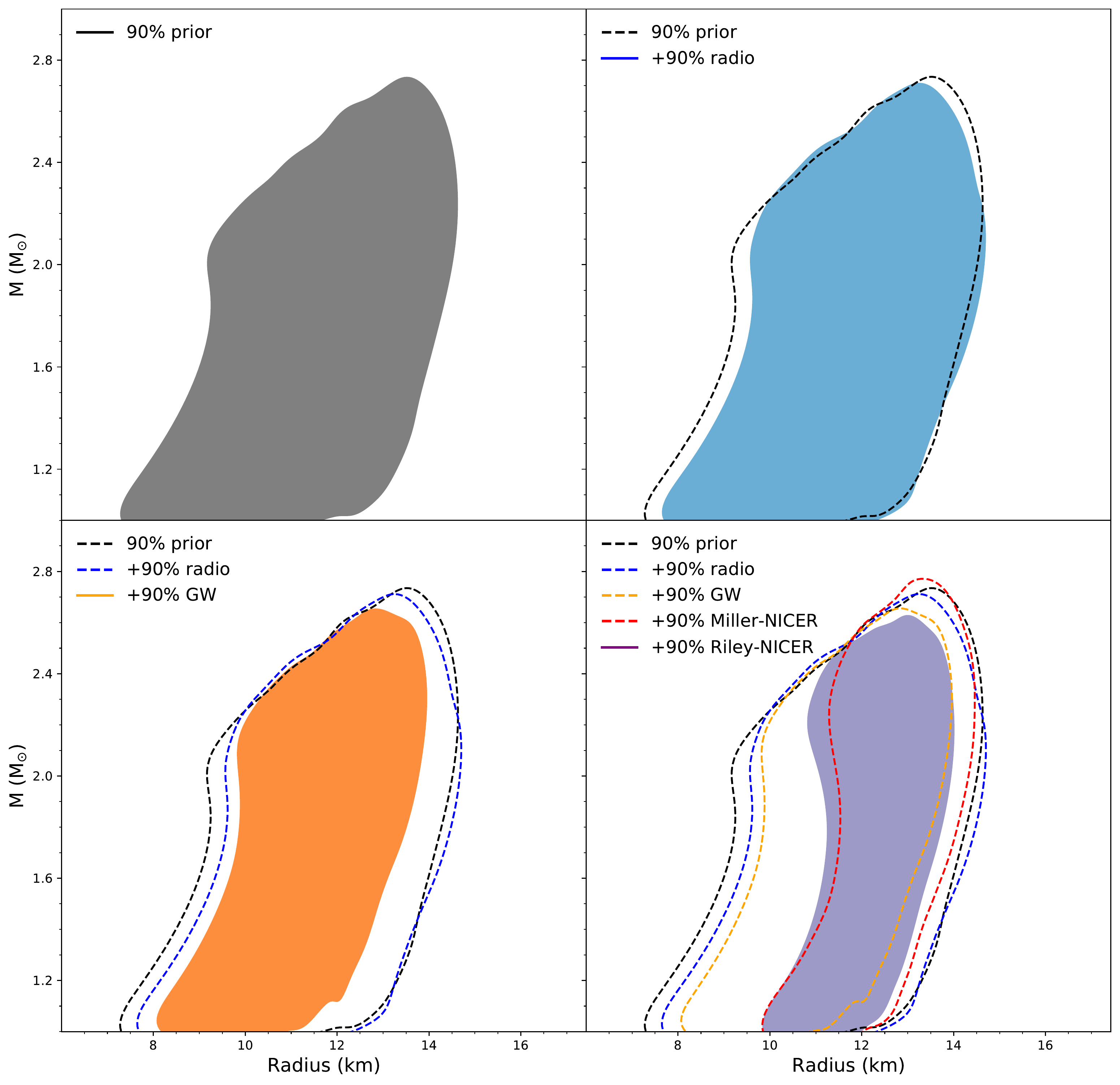}
    \caption{$90 \%$ CI of mass-radius are shown. At first, in the upper left panel prior is shown in black shade. Then upper right panel shows the effect of adding the mass measurement of PSR 0740+6620 from~\cite{Cromartie:2019kug} and compared against the prior. In a similar manner, two BNS merger signals and mass-radius measurements of two pulsars by NICER collaboration are added successively in the lower panel.}
    \label{fig:mr-post}

\end{figure*}

In Fig.~\ref{fig:M-Lambda-fit}, we show the fitted $M-\Lambda$ curves to the true curves for three candidate EoSs SLy (soft)~\cite{Douchin:2001sv}, MPA1 (intermediate)~\cite{MPA1}, and MS1 (stiff)~\cite{MS_PP}. As it can be seen from the bottom panel of this figure the error in this fit is $\leq 5 \%$ for all the considered mass ranges. This parameterization is fitted to wide-ranging candidate EoSs which are listed in Table I of Refs.~\cite{Biswas:2021pvm} using the least square method. We provide the  best-fit parameters in Table~\ref{tab:best-fit}. Based on the values of fitting parameter we choose the prior ranges. In Table~\ref{prior_tab}, the prior ranges of each EoS-sensitive parameters are shown. The boundary values of the prior ranges of $\{b_i\}$ have been fixed by identifying them with the maximum and minimum values in Table \ref{tab:best-fit}.

In order to compute the radius ($R$), we use the well established EoS-insensitive relationship between the compactness ($C=M/R$) and $\Lambda$,

\begin{equation}
\begin{split}
C&=3.531 \times 10^{-1} -2.889 \times 10^{-2} \ln \Lambda \\ &-1.056 \times 10^{-3} (\ln \Lambda)^2 
+1.295 \times 10^{-4} (\ln \Lambda)^3 \,,
\end{split}
\end{equation}
where the coefficients are obtained by fitting all the candidate EoSs considered in this work. We acknowledge there is a maximum of $\sim 3 \%$ error present in this relation for the EoSs considered here. Note, we are using EoS-insensitive $C-\Lambda$ relation only for the computation of the radius. This relation has not been used for the inference of any other physical parameter in our work.

\section{Bayesian methodology}

The posterior of the EoS-sensitive parameters (denoted as $\theta$) are computed using nested sampling algorithm implemented in~{\tt Pymultinest}~\citep{Buchner:2014nha}:
\begin{equation}
    P(\theta | {d}) = \frac{P ({d} | \theta) \times P(\theta)}{P(d)}\, = \frac{\Pi_i P ({d_i} | \theta) \times P(\theta)}{P(d)}\,,
    \label{bayes theorem}
\end{equation}
where $\theta = ( b_0 , b_1, b_2, b_3)$ is the set of our EoS-sensitive parameters, $d = (d_{\rm GW}, d_{\rm X-ray}, d_{\rm Radio})$ is the set of data from the three different 
types of observations that are
used to construct the likelihood, $P(\theta)$ are the priors of those parameters and $P(d)$ is the Bayesian evidence, given the particular EoS model. The expressions of individual likelihood are derived in Refs.~\cite{Landry_2020PhRvD.101l3007L,Biswas_arXiv_2008.01582B,Biswas:2021yge}, but for clarity we repeat them here. For GW observations, information about EoS-sensitive parameters come from the masses $M_1, M_2$ of the two binary components and the corresponding tidal deformabilities $\Lambda_1, \Lambda_2$. In this case, 
\begin{align}
    P(d_{\mathrm{GW}}|\theta) = \int^{M_{\mathrm{max} }}_{M_2}dM_1 \int^{M_1}_{M_{\mathrm{min} }} dM_2 P(M_1,M_2|\theta)   \nonumber \\
    \times P(d_{\mathrm{GW}} | M_1, M_2, \Lambda_1 (M_1,\theta), \Lambda_2 (M_2,\theta)) \,,
    \label{eq:GW-evidence}
\end{align}
where $P(M_1,M_2|\theta)$ is the prior distribution over the component masses which should be informed by the NS population model. However, the choice of wrong population model starts to bias the results significantly only after $\sim 20\mbox{-}30$ observations~\citep{Agathos_2015,Wysocki-2020,Landry_2020PhRvD.101l3007L}. Therefore given the small number of detections at present, this can be fixed by a simple flat distribution over the masses,
\begin{equation}
    P(m|\theta) = \left\{ \begin{matrix} \frac{1}{M_\mathrm{max} - M_\mathrm{min}} & \text{ iff } & M_\mathrm{min} \leq M \leq M_\mathrm{max}, \\ 0 & \text{ else, } & \end{matrix} \right.
\end{equation}
In this work we set $M_\mathrm{min} = 1M_{\odot}$ and $M_\mathrm{max}$ to the maximum mass for that particular EoS. The normalization factor on the NS mass prior is very important as it prefers the EoS with slightly larger $M_\mathrm{max}$ than the heaviest observed NS mass and disfavor EoS with much larger $M_\mathrm{max}$. For example, two EoSs with $M_\mathrm{max}=2.5 M_{\odot}$ and $M_\mathrm{max}=3 M_{\odot}$, will have mass prior probability $P(M|\theta)=2/3$ and $P(M|\theta)=1/2$ respectively if $M_\mathrm{min} \leq M \leq M_\mathrm{max}$. Though both of the EoSs support the heaviest NS mass measurement ($2.08 \pm 0.07 M_{\odot} $) equally well, EoS with $M_\mathrm{max}=3 M_{\odot}$ is less probable than EoS with $M_\mathrm{max}=2.5 M_{\odot}$. Similar approach has been employed in previous works as well~\citep{Miller:2019nzo,Raaijmakers:2019dks,Landry_2020PhRvD.101l3007L,Biswas:2021yge}. Alternatively, the NS mass distribution can be truncated to a largest population mass which is informed a formation channel (eg. supernova)---in that situation EoSs with $M_\mathrm{max}$ greater than the largest population mass will be assigned equal probability. As we lack the knowledge on the upper limit of NS mass distribution, we choose to limit $M_\mathrm{max}$ based on EoS itself not the formation channel. A broader discussion on different choices of mass priors have been discussed in the appendix of Ref.~\citep{Legred:2021hdx}. However, if the masses in GW observation are expected to be smaller than the mass of the heaviest pulsar, then the upper limit on the mass prior can be chosen by the likelihood's domain of support to reduce the computational time.

Equation~\ref{eq:GW-evidence} can further be simplified by fixing the GW chirp mass to its median value with not so much affecting the result~\citep{Raaijmakers:2019dks} given its high precision measurement. As a result we will have one less parameter to integrate over as $M_2$ will be a deterministic function of $M_1$.

From X-ray observations we get the mass and radius measurements of NS. Therefore, the corresponding likelihood takes the following form,
\begin{align}
    P(d_{\rm X-ray}|\theta) = \int^{M_{\mathrm{max} }}_{M_{\mathrm{min} }} dM P(M|\theta) \nonumber \\ \times
    P(d_{\rm X-ray} | M, R (M, \theta)) \,.
\end{align}
Similar to the GW observation, here also the explicit prior normalization over the mass should be taken into account or can be chosen by the Likelihood's domain of support (if applicable). 

 Radio observations provide us with accurate measurements of the NS mass. In this case, we marginalize over the observed mass taking into account its measurement uncertainties,
\begin{align}
    P(d_{\rm Radio}|\theta) = \int^{M_{\mathrm{max} }}_{M_{\mathrm{min} }}dM   P(M|\theta)
    P(d_{\rm Radio} | M) \,.
\end{align}
Here the prior normalization of mass must be taken into account as the observed mass measurement is close to the maximum mass predicted by the EoS.

 The likelihood distributions used in this work are modelled as follows: (a) old mass measurement of PSR J0740+6620~\citep{Cromartie:2019kug} is modelled with a Gaussian likelihood of $2.14 M_{\odot}$ mean and $0.1 M_{\odot}$ $1 \sigma$ standard deviation.  (b). Mass and tidal deformability measurement from GW170817~\cite{Abbott:2018wiz} and GW190425~\cite{Abbott:2020uma} are modelled with an optimized multivariate Gaussian kernel density estimator (KDE) implemented in {\tt Statsmodels}~\cite{seabold2010statsmodels}. (c) Similarly mass and radius measurement of PSR J0030+0451~\cite{Riley:2019yda,Miller:2019cac} and PSR J0740+6620~\cite{Riley:2021pdl,Miller:2021qha} are also modelled with Gaussian KDE. Since the uncertainty in the mass-radius measurement of PSR J0740+6620 is larger for Ref.~\cite{Miller:2021qha} than Ref.~\cite{Riley:2021pdl} due to a conservative treatment of calibration error, we analyze both data separately and compare the results. This is to emphasize when we include the X-ray mass and radius measurement of PSR J0740+6620, we do not add its radio mass measurement to avoid double-counting.

\begin{table*}[ht!]
\begin{tabular}{|c|c|c|c|c|c|c|}

\hline
 Quantity&Prior& Radio & +GW & \multicolumn{3}{c|}{+X-ray}\\
& & & &    +Riley        &  +Miller & Combined  \\
\hline 
$R_{1.4} [\rm km]$ &$11.12_{-2.14}^{+2.48}  $  &$11.19_{-1.92}^{+2.35}  $ & $11.00_{-1.47}^{+1.58}$ &  $11.95_{-0.80}^{+0.96}$&$12.18_{-0.88}^{+0.91}$&$12.05_{-0.87}^{+0.98}$\\

\hline
$\Lambda_{1.4}$ & $228_{-175}^{+559}$  & $237_{-171}^{+529}$ & $212_{-132}^{+280}$ & $359_{-128}^{+218}$ &$405_{-152}^{+224}$&$372_{-150}^{+220}$\\
\hline

$R_{2.08} [\rm km]$ &$11.83_{-2.69}^{+2.74}  $  &$11.92_{-2.24}^{+2.70}  $ & $11.88_{-1.83}^{+1.84}$ &  $12.50_{-1.39}^{+1.35}$&$13.09_{-1.52}^{+1.21}$&$12.65_{-1.46}^{+1.36}$\\

\hline
$M_{\rm max} (M_{\odot})$ & $2.41_{-0.33}^{+0.46}$& $2.40_{-0.29}^{+0.46}$ & $2.41_{-0.28}^{+0.44}$ & $2.47_{-0.27}^{+0.34}$ &$2.57_{-0.31}^{+0.30}$&$2.52_{-0.29}^{+0.33}$\\

\hline 
\end{tabular}
\caption{Median and $90 \%$ CI of $R_{1.4}$, $\Lambda_{1.4}$, $R_{2.08}$, and $M_{\rm max}$ are quoted here.   
}
\label{tab-result}
\end{table*}

\begin{figure*}[ht!]
    \centering
    \includegraphics[width=\textwidth]{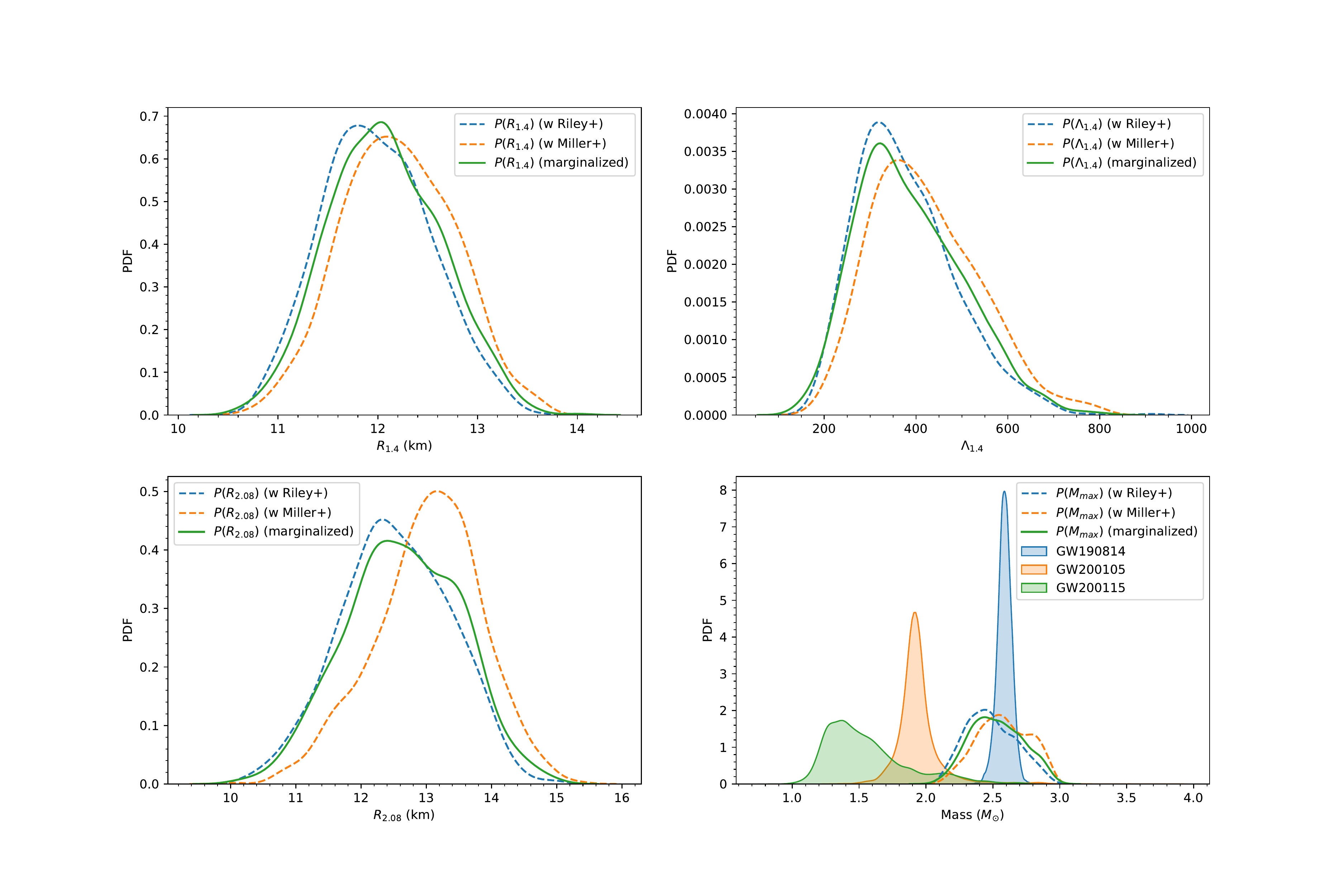}
    \caption{Posterior distributions of $R_{1.4}$, $\Lambda_{1.4}$, $R_{2.08}$, and $M_{\rm max}$ are compared between Riley et al. and Miller et al. dateset and also their marginalized distributions are overlaid. Distribution of the measured secondary masses from three potential NSBH mergers  are shown in the lower panel along with $M_{\rm max}$ distributions.}
    \label{fig:macro_prop}
\end{figure*}

\section{Results}
In Fig.~\ref{fig:post-params}, posterior distribution of EoS-sensitive parameters and their correlations with $R_{1.4}$, $\Lambda_{1.4}$, $R_{2.08}$, and $M_{\rm max}$ are shown after adding successive observations. The corresponding median and $90 \%$ credible interval (CI) of these macroscopic quantities are listed in Table~\ref{tab-result}. We note that the inferred $R$ or $\Lambda$ values only have been sightly larger after adding the old radio mass measurement ($2.14 \pm 0.1 M_{\odot}$) of PSR J0740+6620. All the candidate EoSs which we choose to set the prior range of $b$'s, have $M_{\rm max} \geq 2 M_{\odot}$. As a result, addition of PSR J0740+6620's mass measurement has marginal effect on the $R$ or $\Lambda$ values. This is nicely illustrated in the upper left panel of Fig.~\ref{fig:mr-post}. Here we plot the $90 \%$ CI of mass-radius posterior distribution adding various types of observational data successively. In the upper left panel, the prior region is shown using the solid black . In the upper right panel, the mass-radius posterior region is overlaid after adding PSR J0740+6620's mass measurement. We can see it only exclude a little portion with smaller radii. Thereafter, we add the $M-\Lambda$ measurements from two GW events named GW170817 and GW190425. Addition of GW data makes the values of $R$ and $\Lambda$ much smaller. Within the $90 \%$ CI $R_{1.4}$ and $\Lambda_{1.4}$ now becomes $11.00_{-1.47}^{+1.58}$ km and $212_{-132}^{+280}$ respectively. We note that this bound on $\Lambda_{1.4}$ is in well agreement with the LVC results obtained using EoS-insensitive analysis~\cite{Abbott:2018exr}. Corresponding $90 \%$ CI of mass-radius is overlaid in orange shade at the lower left panel of Fig.~\ref{fig:mr-post}.

Finally, we add the X-ray mass and radius measurement of PSR J0030+0451 and PSR J0740+6620 using the datasets both from Riley et al. and Miller et al. We find the addition of NICER yields larger values of $R$ and $\Lambda$. Interestingly, the posterior of $\widetilde{\Lambda}$ published by the LVC has a bi-modality: The primary mode peaks at $\widetilde{\Lambda} \sim$ 200 and the secondary one peaks at $\widetilde{\Lambda} \sim$ 500. We also see the inferred posterior of $\Lambda_{1.4}$ peaks around $\sim 200$ using the GW data. After adding two NICER observations, we find $\Lambda_{1.4}$ prefers the secondary mode. In the lower right panel of Fig.~\ref{fig:mr-post}, the resulting mass-radius posteriors are compared using both datasets. The datasets from Miller et al. clearly favors slightly larger radii compare to Riley et al. There is $\sim 0.18 (0.45)$ km difference in $R_{1.4} (R_{2.08})$ towards the larger value when we add data from Miller et al. instead of Riley et al.

As there are differences in the inferred posteriors using the two independent datasets, we decide to marginalize over them. One common practice is to take equal number of posterior samples from both analysis and combine them into a single posterior. However giving equal-weighted probability on the both datasets might not be correct way as they correspond to different Bayesian evidence i.e, one being more likely compared to the other. As discussed in Ref.~\cite{Ashton:2019leq}, it is better to weight the samples by their corresponding posterior mixing fraction

\begin{equation}
    \xi_{l} = \frac{Z_{l}}{\sum_{i=0}^{l}Z_{l}}\,,
\end{equation}
where $Z_l$ is the evidence of the $l$-th model. We find the log-evidences ($\ln Z$) computed using Miller et al. and Riley et al. dataset are $-11.47 \pm 0.01$ and $-10.81 \pm 0.02$, respectively. This results into a mixing fraction of $0.34$ for Miller et al. and $0.66$  for Riley et al. dataset. As GW data favor the lower NS radii, Riley et al. dataset is slightly favored compared to Miller et al. In Fig.~\ref{fig:macro_prop}, we overlaid the mixed posterior distributions for the each individual properties along with the Riley et al. and Miller et al. dataset. We find the mixtures are slightly closer to the results obtained using Riley et al. dataset.  

Without the presence of an electromagnetic counterpart and the significant evidence of tidal deformation, the posterior distribution of $M_{\rm max}$ can be used to classify the nature of secondary component in a potential NSBH merger system~\cite{Abbott:2020khf,Essick-2020arXiv,Biswas:2020xna,Tews:2020ylw}. In the third observing run LVC detected three confirmed signals named GW190814~\cite{Abbott:2020khf}, GW200105 and GW200115~\cite{LIGOScientific:2021qlt} whose progenitor might contain one NS. In the lower right panel of  Fig.~\ref{fig:macro_prop}, posterior distribution of the secondary masses are plotted for these three signals and also the $M_{\rm max}$ distributions obtained from our analysis are overlaid. Following Ref.~\cite{Biswas:2020xna}, we compute the probability of the secondary mass being greater than $ M_{\rm max}$ (combined), i.e., $P(M_2 > M_{\rm max})$ = $P(M_2 - M_{\rm max})$, where $M_i$ is the mass of the $i$-th component of the binary. Assuming NS and BH mass distribution do not overlap with each other, we find that the secondary in GW190814, GW200105, and GW200115 respectively has a maximum 62 \%, 2 \%, and 1 \% probability being a BH.

This is to note that the joint detection of GW170817 and its electromagnetic counterparts could provide further constraint on the properties of NS~\cite{Bauswein_2017,Radice:2017lry,Coughlin:2018fis,Capano:2019eae,Breschi_2021}. However, we intentionally do not include that information as these constraints are rather indirect and need careful modeling of the counterparts. The counterpart of GW170817 suggests that the merger remnant collapsed to BH shortly after the merger. Assuming this an upper limit on $M_{\rm max} \lesssim 2.33 M_{\odot}$ has been proposed~\cite{Rezzolla:2017aly}. We do not employ this bound in our analysis and therefore, our constraint on $M_{\rm max}$ extends to the higher values.

\section{Discussion}

In this work, we have constructed a new method to probe the macroscopic properties of NSs and infer the nature of the EoS from them. In the process, we have found a parameterized polynomial structure between $M$ and $\ln\Lambda$. The coefficients of the polynomial depend on the EoSs. Therefore, the measurement of these coefficients can be used to probe the EoS of NS. We fitted this 4-parameter polynomial with the numerically found $M-\Lambda$ relations for multiple EoSs. The resulting fitting error in $M-\Lambda$ curve is allways lesser than $\sim 2.7\%$, except MS1 and MS1B EoS. Even though from all the current observations MS1 family seems to be ruled out~\cite{Biswas:2021pvm}, we still keep them to make our claims as conservative as possible. For the EoSs that do not fall in the MS1 family the fitting error is much lower. We also would like to point out that EoSs with phase transition are not included in this work, but it would be interesting to explore these avenues in the future.

In the parameterized $M-\Lambda$ curve, the dominant contribution on $M$ comes from the zeroth order parameter i.e. $b_0$. From Table~\ref{tab:best-fit}, it can be seen that the value of $b_0$ is very close to the $M_{\rm max}$ i.e., they are correlated. For a larger $M_{\rm max}$, $b_0$ is also larger. This correlation is clearly visible in Fig.~\ref{fig:post-params}. There is a strong correlation (Pearson correlation coefficient (PCC) $\sim 0.99$ ) present between them irrespective of the combination of dataset used in this work. The second most dominant contribution on $M$ comes through the next order parameter i.e. $b_1$. Therefore, it is obvious to have an anti-correlation correlation between $b_0$ and $b_1$. These two parameters can compensate each other to reproduce the same $M$. Interestingly, this correlation is not visible either using radio or radio+GWs data. It has started to appear (though weak, PCC $\sim 0.58$ with Miller et al. data) after adding two NICER observations. This shows the importance of adding multiple observations. As $b_0$ and $M_{\rm max}$ are correlated with each other and $b_0$ and $b_1$ are anti-correlated with each other which directly implies $b_1$ and $M_{\rm max}$ must be anti-correlated. Same as $b_0$ and $b_1$, we see a weak anti-correlation (PCC $\sim 0.48$ with with Miller et al. data) has also started to appear between $b_1$ and $M_{\rm max}$ after adding the two NICER observations. Another significant correlation (PCC $\sim 0.84$ with Miller et al. data) is present between $M_{\rm max}$ and $R_{2.08}$. As $2.08 M_{\odot}$ represents a heavier NS, it is no surprise that $M_{\rm max}$ and $R_{2.08}$ are correlated with each other. As a result, $b_0$ and $R_{2.08}$ are also correlated with PCC $\sim 0.79$ using Miller et al. data. Therefore, we expect $b_0$ will get better constrained if in future we measure high-mass NS radii. 

Since the systematic error (fitting error) is $\lesssim 5\%$, this is quite lower than other approaches. One such approach is to use the universal realtion betwenn $2\Lambda_s \equiv \Lambda_1 + \Lambda_2$ and $2\Lambda_a \equiv \Lambda_1 - \Lambda_2$ \cite{Yagi:2015pkc, Yagi:2016qmr}. In this approach the approximate universality holds to $\sim 20\%$, if $M_1 \lesssim 1.6M_{\odot}$ for all $M_2$. However, from Fig. 4 of Ref. \cite{Yagi:2016qmr} it can be observed that the maximum fractional difference can reach $\sim 50\%$ when $M_1 \sim 2M_{\odot} \sim M_2$. For the binary mass ratio $q = 1$, the maximum fractional difference increases from $\sim10\%$ to $\sim 50\%$ with increasing masses. As a result, this approach has very large systematic errors for equal mass binaries and for NSs with larger masses. In this approach it is also not possible to stack multiple observations.

Another approach is to extract the (mass-independent) coefficients $(c_0, c_1, . . .)$ of a Madhava-Taylor expansion~\cite{Bag1976, Joseph:2011, Phoebe2014} of the tidal deformabilities about some fiducial mass $(M_0)$ \cite{Messenger:2011gi, DelPozzo:2013ala, Yagi:2016qmr}. The goodness of this representation depends on the convergence of the expansion. It was shown in Ref. \cite{Yagi:2016qmr} that for MS1 EoS and $M_0 = 1.4M_{\odot}$, increasing the number of terms in the series decreases the error only in the region $M < 1.9M_{\odot}$. The series diverges in the high mass region $M > 1.9M_{\odot}$. This shows that this parametrization is not suitable for NS binaries with masses that are sufficiently different from $M_0$. By considering multiple EoSs they showed that a certain level of accuracy can be achieved by using up to 3 terms in the series only within a range of $M$, for a given $M_0$. As an example, the fractional difference is smaller than $10\%$ for
$M_0 = 1.4M_{\odot}$ only within $1.1M_{\odot} < M < 1.6M_{\odot}$. If one moves away from this range then the error can increase significantly $\sim 25\%$ or more. Although it is possible to stack multiple observations in this approach but the large systematic error can lead to erroneous conclusions.

Although the current observations have statistical error $\sim 10\%$, it will be at the sub percent level in third generation detectors with multiple detections \cite{Carson:2019rjx}. Hence, it is needed to construct EoS-insensitive approaches that have low systematic error. At the time of writing this paper, as per our knowledge, this has not been achieved. Hence, our approach, where the systematic error is $\lesssim 5\%$ ($\lesssim 2.7\%$ if MS1 family is ignored), has the potential to resolve some of these issues. Since our EoS-insensitive method uses a parameterized function to probe the EoS, it also allows the stacking of multiple observations and finding a joint constraint on EoSs from them. Most importantly, using this approach we avoid solving TOV equation and calculating $\Lambda$ as we directly sample $M, R$, and $\Lambda$ for a given set of $\{b_i\}$. Therefore, it saves a huge computational cost. In comparison to a $P-\rho$ based EoS parameterization used in Refs.~\cite{Biswas_arXiv_2008.01582B,Biswas:2021yge}, it takes a factor of $\sim 10$ less time in the inference. Therefore, with the increase in number of observations, this would certainly be a fast approach to infer the EoS of NSs combining multiple observations.

\section*{Acknowledgement} 
SD thanks Anirban Samaddar for useful discussions.
We gratefully acknowledge the use of high performance super-computing cluster Pegasus at The Inter-University Centre for Astronomy and Astrophysics (IUCAA) for this work.
We would like to thank University Grants Commission (UGC), India, for financial support for a senior research fellowship.
B.B also acknowledges the support from the Knut and Alice Wallenberg Foundation 
under grant Dnr. KAW 2019.0112.

\bibliography{mybiblio}

\end{document}